\documentstyle[12pt]{article}
\def\half{{1\over 2}}

\def\beq{\begin{eqnarray}}
\def\endeq{\end{eqnarray}}
\def\nn{\nonumber \\}
\def\slashed#1{#1\kern-.5em/}

\def\vac{|0>}
\def\N{{\cal N}}
\def\O{{\cal O}}
\def\T{{\cal T}}

\def\A{{\bar A}}
\def\B{{\bar B}}
\def\C{{\bar C}}
\def\g{{\tilde g}}
\def\mes#1{{dq_#1\over \sqrt{2 \pi q_#1}}}
\def\Romnum#1{\uppercase\expandafter{\romannumeral #1}}

\makeatletter
\ifx\@revmess\undefined
\renewcommand{\baselinestretch}{2}\tiny\normalsize
\def\theequation{\arabic{section}.\arabic{equation}}
\def\preprint#1{%
\noindent\hfill\hbox{#1}\vskip -6pt
}

\def\address#1{{\small\it\centering\ignorespaces#1\vskip.5in}}
\def\pacs#1{\par\noindent #1}
\def\acknowledgements{\bigskip\centerline{ACKNOWLEDGEMENT}\bigskip}
\def\maketitle{}
\def\title#1{{\par \Large\rm\centering #1\vskip.5in}}
\def\author#1{{\par\centering #1\par}}
\def\thesection{\Roman{section}. }
\makeatletter
\@addtoreset{equation}{section}
\makeatother
\def\appendix{
\setcounter{section}{0}
\setcounter{equation}{0}
\def\thesection{APPENDIX. }
\def\theequation{\Alph{section}.\arabic{equation}}
}

\makeatletter
\long\def\@makefntext#1{%
\renewcommand{\baselinestretch}{1}\tiny\normalsize%
\parindent 1em\noindent
\hbox to 2em{\hss$^{\@thefnmark}$~}%
\@tempdima\columnwidth\advance\@tempdima-2em\parbox[t]%
{\@tempdima}{#1}%
\renewcommand{\baselinestretch}{2}\tiny\normalsize
}
\makeatother
\def\section#1{%
\addtocounter{section}{1}%
\setcounter{equation}{0}
\par\bigskip\centerline{\large\rm\thesection #1}\bigskip\indent}
\def\thebibliography#1{\list
{[\arabic{enumi}]}{\settowidth\labelwidth{[#1]}\leftmargin\labelwidth
\advance\leftmargin\labelsep
\usecounter{enumi}}
\def\newblock{\hskip .11em plus .33em minus .07em}
\sloppy\clubpenalty4000\widowpenalty4000
\sfcode`\.=1000\relax}
\newenvironment{references}%
{\clearpage\center{\rm \large REFERENCES}\bigskip
\begin{thebibliography}{99}}%
{\end{thebibliography}}
\fi
\makeatother
\begin{document}
\preprint{OHSTPY-HEP-T-96-032}
\preprint{DOE/ER/01545-699}
\preprint{HUEAP-010}
\title{Mesons in 2-Dimensional QCD on the Light Cone}
\author{
Osamu Abe\footnote{Permanent address: Physics Laboratory, 
Asahikawa Campus, Hokkaido University of Education, 9 Hokumoncho, 
Asahikawa 070, Japan}\footnote{e-mail address: 
osamu@asa.hokkyodai.ac.jp}, 
Gordon J. Aubrecht \Romnum{2}\footnote{e-mail address: 
aubrecht@mps.ohio-state.edu}, and 
K. Tanaka\footnote{e-mail address: tanaka@mps.ohio-state.edu} 
}
\address{
Department of Physics\\ The Ohio State University\\ 
174 West 18th Avenue\\ Columbus, Ohio 43210 
}
\date{\today}
\maketitle
\begin{abstract}
Two dimensional QCD is quantized on the light front coordinate. We 
solve the Einstein-Schr\"odinger equation by the use of Tamm-Dancoff 
truncation and find that the simplest wavefunction produces the $M/g$ 
versus $m/g$ relation in agreement with other calculations, where $M$ 
and $m$ are the masses of the ground state and quarks, respectively.  
\end{abstract}
\pacs{11.10.Kk, 11.15.Pg, 11.15.Tk}
\newpage \section{INTRODUCTION}
   The light front Tamm-Dancoff (LFTD) method \cite{wilson} has been 
introduced as an alternative tool to lattice gauge theory to 
investigate relativistic bound states nonperturbatively. The LFTD is 
a Tamm-Dancoff (T-D) approximation \cite{tamm} applied to field 
theories quantized on the light front coordinate. In the usual 
perturbative field theory quantized in the equal time frame, a vacuum 
state is an infinite sea of constituents such as electrons and 
photons in QED and quarks and gluons in QCD. Bound states of a 
hydrogen atom in QED, and mesons and baryons in QCD, arise as 
excitations of this sea.\par 
   The T-D method independently considered the possibility of 
describing the vacuum and the bound states with a finite number of 
particles, and solving a set of coupled integral equations. This 
method was applied to a variety of problems in strong interactions, 
but it was unsuccessful because a large number of amplitudes was 
required to solve a given problem.\par 
   On the other hand, in the LFTD method all constituents have 
non-negative longitudinal momenta defined by $p^+ = {p^0 + p^3 \over 
\sqrt 2}$ in the light front coordinate. Then the vacuum of the 
system under consideration can not have constituents, or the physical 
vacuum is equivalent to the bare vacuum in the light front 
coordinate. Therefore, we may expect this approach to remove a 
serious problem in the T-D approximation in the equal time frame.\par 
   It was found in several problems such as the hydrogen atom, 
positronium, and the two-dimensional Yukawa interaction, that one 
needed an additional T-D amplitude to obtain the particle spectrum 
\cite{perry}. We have provided an argument \cite{abe1} for why this 
happens to be the case.\par 
   In 1974, 't Hooft \cite{thooft} introduced a model, 2 dimensional 
QCD, that continues to be studied. We examine this model in the 
framework of the LFTD method. 
\vfill\eject
\section{MESONS IN 2D QCD}
   We consider the quantized version of Einstein's equation 
\begin{equation} 2P(H_0 + H_I)| \Psi > = M^2 | \Psi >, 
\label{einstein}
\end{equation} 
in two dimensional QCD with $SU(N_c)$ symmetry, for states with mass 
$M$ on the light front coordinate. Here $P$ is the momentum operator 
in the $SU(N_c)$ gauge, $H_0$ is the free part of the Hamiltonian, 
given by $H_0\equiv P^-_{free}+P^-_{self}$, and $H_I=P^-_0+P^-_2$ is 
the interaction Hamiltonian. We consider a meson state $|\Psi(f,{\bar 
f}';p) >$ with momentum $p$, which is described by 
\beq
& &|\Psi(f,{\bar f}';p) > = \nn 
& &\sum_c \int_0^\infty
\mes{1}\mes{2}\delta (p-q_1-q_2) \psi_2(f,{\bar f}';q_1,q_2)
b^{\dag}(f,q_1,c) d^{\dag}(f',q_2,c) \vac\nn 
&+&\sum_{f_1,f_2,f_3,f_4}\sum_{c_1,c_2,c_3,c_4}\int_0^\infty
\mes{1}\mes{2}\mes{3} \mes{4}\delta (p - q_1- q_2 -q_3 - q_4) \nn 
&\times&
\psi_4(f,{\bar f}';f_1,f_2,{\bar f}_3,{\bar
f}_4;q_1,q_2,q_3,q_4;c_1,c_2,c_3,c_4)\nn  
&\times&
b^{\dag}(f_1,q_1,c_1)b^{\dag}(f_2,q_2,c_2) d^{\dag}(f_3,q_3,c_3) 
d^{\dag}(f_4,q_4,c_4) \vac . \label{mesonstate} 
\endeq
The first order LFTD approximation results from putting 
Eq.~(\ref{mesonstate}) into Eq.~(\ref{einstein}) and projecting the 
resultant equation onto a state with a fermion and an anti-fermion, 
and a state with two fermions and two anti-fermions. This leads to 
\beq
<f_1,q_1,c_1;{\bar f}_2,q_2,c_2| \left( M^2 - 2P H_0\right) |\Psi> = 
<f_1,q_1,c_1;{\bar f}_2,q_2,c_2| 2P H_I|\Psi>, \label{two} 
\endeq
and 
\beq
& &<f_1,q_1,c_1;f_2,q_2,c_2;{\bar f}_3,q_3,c_3;{\bar f}_4,q_4,c_4| 
\left(
M^2 - 2P H_0\right) |\Psi> \nn  
&=& <f_1,q_1,c_1;f_2,q_2,c_2;{\bar
f}_3,q_3,c_3;{\bar f}_4,q_4,c_4| 2P H_I|\Psi>, \label{four} 
\endeq
where 
\beq
|f_1,q_1,c_1;{\bar f}_2,q_2,c_2>={b^{\dag}(f_1,q_1,c_1) \over 
\sqrt{2\pi q_1}} {d^{\dag}(f_2,q_2,c_2) \over \sqrt{2\pi q_2}}\vac , 
\endeq
and 
\beq
\lefteqn{|f_1,q_1,c_1;f_2,q_2,c_2;{\bar f}_3,q_3,c_3;{\bar
f}_4,q_4,c_4>}\nn  
&=&{b^{\dag}(f_1,q_1,c_1) \over \sqrt{2\pi q_1}}
{b^{\dag}(f_2,q_2,c_2) \over \sqrt{2\pi q_2}} {d^{\dag}(f_3,q_3,c_3) 
\over
\sqrt{2\pi q_3}} {d^{\dag}(f_4,q_4,c_4) \over \sqrt{2\pi q_4}}\vac .
\endeq
\par
   It is straightforward to derive coupled integral equations for the 
amplitudes $\psi_2$ and $\psi_4$. The results are 
\beq
& &\delta_{f_1f}\delta_{f_2f'}\delta_{c_1c_2} F(M;f,f';q_1,q_2)
\psi_2(f,{\bar f}';q_1,q_2) \nn
&=&-\delta_{f_1f}\delta_{f_2f'}\delta_{c_1c_2}{N_c^2-1\over N_c} 
{pg^2\over
2\pi}\wp\!\!\int_{-\infty}^{\infty}{dk\over k^2} \psi_2(f,{\bar 
f}';q_1+k,q_2-k) \nn  
&-&{2pg^2\over
2\pi}\sum_{j:flavors}\sum_{d_1,d_2,d_3:colors} 
\T_{c_1d_2;d_3d_1}\wp\!\! 
\int_{-\infty}^{\infty}{dk\over k^2} \int_0^\infty dq\nn 
&\times&
\psi_4(f,{\bar f}';[j,f_1],[{\bar j},{\bar f}_2];
[q,q_1+k],[-q-k,q_2];[d_1,d_2],[d_3,c_2]) \nn  
&-&{2pg^2\over
2\pi}\sum_{j:flavors}\sum_{d_1,d_2,d_3:colors} 
\T_{d_3d_1;d_2c_2}\wp\!\! 
\int_{-\infty}^{\infty}{dk\over k^2} \int_0^\infty dq\nn
&\times&\psi_4(f,{\bar f}';[j,f_1],[{\bar f}_2,{\bar j}];
[q,q_1],[q_2-k,-q+k];[d_1,c_1],[d_2,d_3]), \label{two2} 
\endeq
where $\wp$ stands for principal value integral, $N_c$ is the order 
of the gauge group, and we have used 
\beq
{\cal T}_{cd;ef}&=&{1\over 2}\left\{\delta_{cf}\delta_{de} -{1\over 
N_c}
\delta_{cd}\delta_{ef}\right\},
\endeq
\beq
\lefteqn{F(M;f_1,f_2\cdots ;q_1,q_2\cdots ) }\nn 
&=&M^2-2(q_1+q_2+\cdots )
\left\{{m_{f_1}^2+\alpha (q_1) \over 2q_1} +{m_{f_2}^2+\alpha (q_2) 
\over
2q_2} +\cdots \right\},\label{three2} 
\\ \lefteqn{\psi_4(f,{\bar f}';[f_1,f_2],[{\bar
f}_3,{\bar f}_4];[q_1,q_2], [q_3,q_4];[c_1,c_2],[c_3,c_4])}\nn 
&=&\psi_4(f,{\bar f}';f_1,f_2,{\bar f}_3,{\bar f}_4;q_1,q_2,q_3,q_4;
c_1,c_2,c_3,c_4) -(1\leftrightarrow 2) -(3\leftrightarrow 4)\nn  
&+&
(1\leftrightarrow 2,\; 3\leftrightarrow 4), 
\endeq
and 
\beq
& &F(M;f_1,f_2,f_3,f_4;q_1,q_2,q_3,q_4) \psi_4(f,{\bar f}';[f_1,f_2],
[{\bar f}_3,{\bar f}_4];[q_1,q_2],[q_3,q_4]; [c_1,c_2],[c_3,c_4]) \nn 
&=&{2pg^2\over 2\pi}\Bigl[\Bigl\{
\delta_{f_1f}\delta_{f_2f_3}\delta_{f_4f'}\T_{c_1c_4;c_2c_3} \psi_2(f,
{\bar
f}';q_1+q_2+q_3,q_4) \cdot {1\over (q_2+q_3) ^2}\nn  
&-&(1\leftrightarrow 2)
-(3\leftrightarrow 4) + (1\leftrightarrow 2,\; 3\leftrightarrow 4) 
\Bigr\}\nn 
&-& \Bigl\{
\delta_{f_1f_4}\delta_{f_2f}\delta_{f_3f'}\T_{c_2c_3;c_1c_4} \psi_2(f,
{\bar
f}';q_2,q_1+q_3+q_4) \cdot {1\over (q_1+q_4) ^2}\nn  
&-&(1\leftrightarrow 2)
-(3\leftrightarrow 4) + (1\leftrightarrow 2,\; 3\leftrightarrow 4) 
\Bigr\}\Bigr]. \label{four2}
\endeq
\par
   The $\alpha$ term in Eq.~(\ref{three2}) denotes the meson self 
energy and is found to be equal to \\ $\displaystyle -(N_c^2 
-1)g^2/(2N_c\pi )$.\par 
   Hereafter, we restrict ourselves to the one flavor case. We define 
\beq
p=q_1+q_2,\quad q_1=\left({1\over 2}+x\right)p,\quad \psi_2(q_1,
q_2)\equiv b_0(x),\quad a=\half +x\quad{\rm and}\quad b=\half-x. 
\endeq
Substituting Eq.~(\ref{four2}) into Eq.~(\ref{two2}), we have 
\beq
\Bigl[ M^2- {4(m^2+\alpha ) \over 1-4x^2}\Bigr] b_0(x)
=I_0(x)+I_1(x)+I_2(x), \label{fulleq} 
\endeq
where 
\beq
I_0(x)&\equiv&-2\N_1\wp\int_{-a}^{b}{dy\over y^2}b_0(x+y),
\label{i0}\\ I_1(x)&\equiv&\N_2\wp\int_{-a}^0{dy\over 
y^2}\int_0^{-y}\!\!dz {1 \over F(M;z,a+y,-z-y,b) } \times\nn  
& &\Bigl[\Bigl\{ {n\over (a-z) ^2}+{1\over
y^2}\Bigr\}b_0(x) +\Bigl\{{n\over (b+z) ^2}+{1\over (1+y) 
^2}\Bigr\}b_0(y+z+\half ) \nn  
&-&\Bigl\{{n\over (b+z) ^2}+{1\over
y^2}\Bigr\}b_0(x+y) -\Bigl\{{n\over (a-z) ^2}+{1\over (1+y) 
^2}\Bigr\}b_0(z-\half ) \Bigr], \label{i1} 
\endeq
and 
\beq
I_2(x)&\equiv& -\N_2\wp\int_0^{b}{dy\over y^2}\int_0^y\!\!dz {1 \over 
F(M;-z+y,a,b-y,z) }\nn  
& & \Bigl[\Bigl\{ {n\over (b-z) ^2}+{1\over (1-y)
^2}\Bigr\}b_0(\half -z) +\Bigl\{{n\over (a+z) ^2}+{1\over 
y^2}\Bigr\}b_0(x+y) \nn  
&-&\Bigl\{{n\over (b-z) ^2}+{1\over
y^2}\Bigr\}b_0(x) -\Bigl\{{n\over (a+z) ^2}+{1\over (1-y) 
^2}\Bigr\}b_0(-\half +y-z) \Bigr]. \label{i2} 
\endeq
Here we use the shorthand notation 
\beq
\N_1=(N_c^2-1)g^2/(4N_c\pi),\quad \N_2=(N_c-1)g^4/(4N_c\pi^2), \quad 
{\rm
and}\quad n=(N_c+1)/N_c .\nn 
\endeq
With the definition of Eqs.~(\ref{i1}) and (\ref{i2}) we can show 
\beq
I_2(-x)=\pm I_1(x), \label{sym} 
\endeq
corresponding to the symmetry property of $b_0 (\pm x) = \pm b_0 
(x)$. It is not possible to solve Eq.~(\ref{fulleq}) analytically. 
One usually selects a wavefunction $b_0(x)$ and tries to match both 
sides of Eq. (\ref{fulleq}). It happens that the results do not 
depend sensitively on the form of $b_0(x)$. In other words, different 
forms of $b_0(x)$ lead to similar predictions. Further, the 
contributions of the integrals $I_1(x)$ and $I_2(x)$ are at most 10\% 
of that of $I_0(x)$. Suppose we take the simple form of power series 
expansion for $b_0$. Since we wish to determine the ground state 
wavefunction, which has no node (and thus no odd powers of x), we 
write 
\beq
b_0(x)=1+{\A} x^2 + {\B}x^4 +{\C}x^6+\cdots . \label{power} 
\endeq
\par
   The form of Eq.~(\ref{power}) is motivated by our assumption that 
$b_0(x)$ is an even function and we normalize the first term to 1. The
power series is expressible in terms of a linear combination of Airy 
finctions that are relevant to linear potentials. It is surprising, 
as we shall show, that most of the information is present in the 
first terms $1+ 
\A x^2$ of (\ref{power}) and that higher order terms such as the $\B$ 
and
$\C$ terms
provide a small correction. We expect $\A$ to be negative in order 
that 
$b_0(x)$ decrease away from $x= 0$. The wavefunction $1 - |\A|x^2$ is 
a
maximum at $x=0$ and is zero at $x =\pm |\A|^{-1/2}$.\par 
   As a warmup, consider the simplest model \cite{abe2} of 
(\ref{fulleq}) with only $I_0(x)$ and 
\beq
b_0(x) = 1+ {\A}x^2. \label{simplest} 
\endeq
We study this model because it contains the ingredients of the 
resulting relation of the ground state mass $M$ and the quark mass 
$m$. The additional contributions of $I_1(x)$, $I_2(x)$ and higher 
order terms such as $\B$ and $\C$ terms provide a small correction to 
this relation.\par 
   Define 
\begin{equation} \tilde M^2 = M^2/2N_1 \qquad \tilde {m}^2 =
({m^2 \over 2N_1})-1, 
\end{equation}
where the relation $\alpha = -2N_1$ is used, and write the simplified 
(\ref{fulleq}) as 
\begin{equation} \left\{ (1-4x^2) \tilde M^2 - 4 \tilde {m}^2\right\}
(1+ {\A}x^2) = 4 +{\A} \left\{ 4x^2 -2(1-4x^2)x\ln {1 - 2x \over 1 + 
2x}-1 + 4x^2 \right\}.  
\end{equation}
From equating the coefficients of the $x^0$ and $x^2$ terms, we 
obtain 
\begin{equation} \tilde M^2 - 4 \tilde{m}^2 = 4- \A, \label{x0}
\end{equation}
\begin{equation} 4 \tilde M^2 - ( \tilde M^2 - 4
\tilde{m}^2)\A = -16 {\A}\label{x2}
\end{equation}\par
   We find that when $b_0 (x) = 1$, Eqs.~(\ref{x0}) and (\ref{x2}) 
yield 
$M=m=0$, a relation we regard as a boundary condition. When we 
eliminate
$\A$ from (\ref{x0}) and (\ref{x2}), we obtain
\begin{equation} M/g = \sqrt6\, m/g = 2.45\, m/g \label{simplestMm}
\end{equation} 
and the condition 
$M^2$ be real is $m/g \leq 2 / \sqrt {3 \pi}$. Also, one can express
$\tilde M^2$ and $\tilde {m}^2$ in terms of $\A$,
\begin{equation} {M^2
\over m^2} = {3 + \A/4 \over 1/2 + \A / 16}. \label{Moverm}
\end{equation}\par
  We return to $I_0(x)$ and $I_1(x)$ given by Eqs.~(2.14) and (2.15), 
respectively. The $I_0(x)$ is written in the form 
\beq
I_0(x)&=& {8\N_1\over 1-4x^2} +2{\A}\N_1\left\{{4x^2\over 
1-4x^2}-2x\log{1-2x\over 1+2x} -1\right\}\nn  
&+&2{\B}\N_1\left\{{4x^4\over
1-4x^2}-4x^3\log{1-2x\over 1+2x} -{1\over 12} -3x^2\right\}\nn 
&+&2{\C}\N_1\left\{{4x^6\over 1-4x^2}-6x^5\log{1-2x\over 1+2x} 
-{1\over 80}
-{x^2\over 4}-5x^4\right\} +\cdots , \label{i02} 
\endeq
and 
\beq
I_1(x)={\A}I_A(x)+{\B}I_B(x)+{\C}I_C(x)+\cdots. 
\label{i1iaibic}
\endeq
Here 
\beq
I_A(x)&=&\N_2\int_{-a}^0dy {E(x,y)\over y^2}\int_0^{-y}dz\times\nn  
& &\left\{1+{C(x,y)\over z(z+y)-C(x,y)}\right\}\left[2n\left\{{x\over 
a-z}
+{x+y\over b+z}\right\}-{2x\over y}+{2z-1\over 1+y}\right], 
\label{ia} 
\endeq
\beq
I_B(x)&=&\N_2\wp\int_{-a}^0dy{E(x,y)\over y^2}\int_0^{-y}dz 
\left[1+{C(x,y)\over z(z+y)-C(x,y)}\right]\times\nn 
& &\left[
n\left\{-{z-b\over z-a}\left(x^2+\left(z-\half\right)^2\right) 
+{2y+z+a\over z+b}\left(\left(y+z+\half \right)^2+\bigl( 
x+y\bigr)^2\right) 
\right\}\right.\nn 
& &\left.-{1\over y}(2x+y)\left(x^2+\bigl(
x+y\bigr)^2\right) +{1\over 1+y}(y+2z)\left(\left(y+z+\half\right)^2+ 
\left(z-\half\right)^2\right)\right],\nn 
& & \label{ib}
\endeq
and 
\beq
I_C(x)&=&\N_2\wp\int_{-a}^0dy{E(x,y)\over y^2}\int_0^{-y}dz 
\left[1+{C(x,y)\over z(z+y)-C(x,y)}\right]\times\nn 
& &\left[
n\left\{-{z-b\over z-a}\left(x^4+x^2\left(z-\half\right)^2 
+\left(z-\half\right)^4\right)\right.\right.\nn  
& &\left.+{2y+z+a\over
z+b}\left(\left(y+z+\half \right)^4 +\left(y+z+\half \right)^2\bigl( 
x+y\bigr)^2+\bigl( x+y\bigr)^4\right) \right\}\nn  
& &\left.-{1\over
y}(2x+y)\left(x^4+x^2\bigl( x+y\bigr)^2 +\bigl( 
x+y\bigr)^4\right)\right.\nn  
& &\left.+{1\over
1+y}(y+2z)\left(\left(y+z+\half\right)^4+ 
\left(y+z+\half\right)^2\left(z-\half\right)^2+
\left(z-\half\right)^4\right)\right]\nn 
& &\label{ic}
\endeq
where the functions $E$ and $C$ are defined as 
\beq
E(x,y)={b(a+y)\over M^2 b(a+y)-(m^2+\alpha )(1+y)}, 
\endeq
and 
\beq
C(x,y)={(m^2+\alpha )(a+y)b\,y\over M^2 b(a+y)-(m^2+\alpha )(1+y)}. 
\endeq
We may fit the integrands of $I_A(x)$, $I_B(x)$, and $I_C(x)$ 
empirically after performing the $z$-integral: 
\beq
{\rm Integrand\quad of}\quad I_{A,B,C}(x)&=& f_{1A,B,
C}(y;r)\left(x+{y\over
2}\right) +f_{3A,B,C}(y;r)\left(x+{y\over 2}\right)^3, \label{ia2} 
\endeq
   We return to $I_0(x)$ and $I_1(x)$ given by Eqs.~(\ref{i0}) and 
(\ref{i1}), respectively, that may be written in the form where $r 
\equiv M^2 /( m^2 + \alpha )$ and details of the functions $f_1$ and 
$f_3$ are given in the Appendix. This procedure produces equations 
that may be solved sequentially at each order of $x$ for the unknown 
parameters $\A$, $\B$, 
$\C$, $\cdots$, $r$, and $\g ={g^2\over m^2+\alpha}$,
\beq
\O (x^0)\,&:&\, r-4=-{4\g\over 3\pi}\Biggl[ -4+\A+{\B\over 
12}+{\C\over 80}
\Biggr]\nn 
&+&{\g^2\over 6\pi^2}\Biggl[\A F_{A0}+\B F_{B0}+\C
F_{C0}\Biggr], \label{eqo0} \\ \O (x^2)\,&:&\, (r-4)\A -4r=-{4\g\over 
3\pi}\Biggl[ -16\A+{8\B\over 3} +{\C\over 5}\Biggr]\nn  
&+&{\g^2\over
6\pi^2}\Biggl[\A (F_{A2}-4F_{A0}) +\B (F_{B2}-4F_{B0})+\C 
(F_{C2}-4F_{C0})\Biggr], \label{eqo2}\\ \O(x^4)\,&:&\, (r-4)\B -4r\A 
=-{4\g\over 3\pi}\Biggl[ {64\A\over 3}-32\B+4\C]\nn  
&+&{\g^2\over
6\pi^2}\Biggl[\A (F_{A4}-4F_{A2}) +\B (F_{B4}-4F_{B2})+\C 
(F_{C4}-4F_{C2})\Biggr], \label{eqo4}\\ \O (x^6)\,&:&\, (r-4)\C -4r\B 
=-{4\g\over 3\pi}\Biggl[ {256\A\over 15} +{128\B\over 3} 
-48\C\Biggr]\nn 
&+&{\g^2\over 6\pi^2}\Biggl[\A (F_{A6}-4F_{A4}) +\B 
(F_{B6}-4F_{B4})+\C
(F_{C6}-4F_{C4})\Biggr]. \label{eqo6} 
\endeq
To solve to order $x^2$, we neglect the terms in the expansion of the 
wave function to higher order that involve $\B$, $\C$, etc. This 
yields two equations from (\ref{eqo0}) and (\ref{eqo2}) in terms of 
$\A$, $r$, and 
$\g$. The solution to order $x^4$ involves neglecting terms $\C$, 
etc., in
the wavefunction and using (\ref{eqo0})-(\ref{eqo4}), and so on.\par 
   We can solve Eqs.~(\ref{eqo0})-(\ref{eqo6}) numerically for $r$ in 
terms of $\g$. We show the result in Fig.1. For small values of $m/g$,
 we shall show that in the present model $M/g$ approximately 
satisfies Eq. (\ref{simplestMm}).\par 

\section{VALIDITY OF THE APPROXIMATION.}
   In this section we consider the contributions of $I_1(x)$ and 
$I_2(x)$ or alternatively $I_A(x)$ (\ref{ia}), $I_B(x)$ (\ref{ib}), 
and $I_C(x)$ (\ref{ic}) to the ground state mass $M$. \par 
   At first, we neglect $I_1(x)$ and $I_2(x)$ in Eq.~(\ref{fulleq}). 
We can solve the above equations (\ref{eqo0})-(\ref{eqo6}) 
numerically for $r$ in terms of $\g$ when we neglect these 
contributions of the integrals 
$I_A(x)$, $I_B(x)$, and $I_C(x)$, $\cdots$. Multiplying 
Eq.~(\ref{fulleq})
by $1 - 4x^2$ and setting $\bar D= \dots = 0$ for simplicity, we have 
\beq
\O (x^0)\,&:&\, r-4=-{4\g\over 3\pi}\Biggl[ -4+\A+{\B\over 12} 
+{\C\over
80}\Biggr], \label{eqo00}\\ \O (x^2)\,&:&\, (r-4)\A -4r=-{4\g\over 
3\pi}\Biggl[ -16\A+{8\B\over 3} +{\C\over 5}\Biggr], \label{eqo02}\\ 
\O (x^4)\,&:&\, (r-4)\B -4r\A =-{4\g\over 3\pi}\Biggl[ {64\A\over 
3}-32\B+4\C 
\Biggr],\label{eqo04}\\ \O (x^6)\,&:&\, (r-4)\C -4r\B =-{4\g\over
3\pi}\Biggl[ {256\A\over 15} +{128\B\over 
3}-48\C\Biggr].\label{eqo06} 
\endeq
Then (\ref{eqo00}) to (\ref{eqo02}) with $C$ = 0, after eliminating 
$r$ and 
$\g$, yield
\beq
12{\A^3} + 144{\A^2} + 256\A - 44\A\B + {\A^2}\B - 336\B - {\B^2}=0. 
\label{eqAandB}
\endeq
Solving the above equation for $\A$, we obtain three roots $\A_1(\B)$,
 
$\A_2(\B)$, and $\A_3(\B)$. The solutions where $\A$ and $\B$ are 
real is
shown in Fig. 2. The requirement that $M$ is real rules out the 
solution 
$\A_3$ in Fig. 2.\par
   We can express $M^2$ and $m^2$ in terms of $\A$, $\B$ with the aid 
of (\ref{eqo00}), (\ref{eqo02}) and (\ref{eqo04}): 
\beq
{M^2\over m^2} = \left(3 + {\A \over 4} - {2 \over 3} {\B \over \A} + 
{\B
\over 48} \right) \bigg/ \left({1 \over 2} + {\A \over 16} - { 3 \B 
\over
16 \A} + {\B \over 192}\right). 
\endeq
\par
   We may also repeat this calculation, by including (\ref{eqo06}), 
to find 
$\C$. \par
   Our range of $m/g$ is limited to $0 - 2/\sqrt {3 \pi}$. We have 
included these results in Fig. 1. For small values of $m/g$, $M/g$ 
approximately satisfies (\ref{simplestMm}) (dashed, lowest line).\par 
   We may repeat the procedures given above, but including the 
integrals 
$I_A(x)$, $I_B(x)$, and $I_C(x)$. The changes from the solution of 
Sec. 3
and those calculated immediately above are under $10\%$ for all the 
cases. Overall, Fig. 1 shows that $M/g$ vs. $m/g$ differs at most by 
less than 
$20\%$ from the values obtained from the simplest model.\par
\section{COMPARISON TO OTHER PARAMETRIZATIONS.}
   We compare the relation $M/m$ when different wavefunctions are 
used for 
$b_0(x)$. We begin by writing 't Hooft's equation for a two particle 
state
\cite{thooft} in the present notation,
\beq
{\pi M^2\over g^2}b_0(x) = \left({1\over 1/2 - x}+{1\over 1/2 + x} 
\right)
\left({\pi m^2\over g^2}-1 \right) b_0(x) - 
\int_{-\half}^{\half}dy{b_0(y)
\over (y - x)^2}. \label{thoofteq}
\endeq
Eq.~(\ref{thoofteq}) corresponds to our Eq.~(\ref{fulleq}). Both 
sides of Eq.~(\ref{thoofteq}) are integrated with respect to $x$, and 
we obtain 
\beq
M^2\int_{-\half}^\half b_0(x) dx =4 m^2 \int_{-\half}^\half dx 
{b_0(x) 
\over (1 - 4x^2)}. \label{sugiharaeq}
\endeq
When we substitute 't Hooft's wavefunction $\sin (n\pi (x + 1/2)) = 
\cos 
\left(n\pi x + {(n-1)\pi \over 2}\right)$ in
(\ref{i1iaibic}), we find by inspection that only the odd numbers of 
$n$ contribute to Eq.~(\ref{sugiharaeq}) and obtain 
\beq
{2 M^2/n\pi} = 2 m^2 {\rm Si}(n\pi),\quad n = 1, 3, 5,\cdots ,
\endeq
where the sine integral ${\rm Si}(z)$ is 
\begin{equation} {\rm Si}(z) =\int_0^z dt {\sin t\over t}. 
\end{equation}
For $n = 1$ 
\beq
M^2 =\pi {\rm Si}(\pi) m^2 =5.82\, m^2 ,\quad {\rm or}\quad 
{M/g}=2.41\,{m/g}. 
\label{thooftresult}
\endeq
which is in good agreement with (\ref{simplestMm}).\par 
   If a wavefunctions of the form \cite{sugihara} $b_0(x) = 
(1-4x^2)^\beta$ is used, where $\beta > 0$ and for the range of small 
$\beta$, we obtain $b_0(x) \approx 1 - 4\beta x^2$. Upon comparison 
with
(\ref{simplest}) we get $\A = -4 \beta$. One has for such 
wavefunctions 
$(1-4x^2)^\beta$ the boundary condition \cite{thooft,sugihara}
\beq
{2\pi m^2\over g^2 \left(N_c-{1\over N_c}\right)} -1+\pi\beta \cot
(\pi\beta )=0. \label{thoofteq2} 
\endeq
In the limit $N_c\rightarrow\infty$, we get $\tan \pi\beta =\pi\beta$,
 which holds when $\beta$ is small.\par 
   For small $\beta$ and $N_c=3$, Eq.~(\ref{thoofteq2}) is given by 
\beq
3 \pi m^2/4g^2 - \pi^2 \beta^2 (1+3\pi m^2/8g^2)/ 3 = 0, 
\endeq
so that 
\begin{equation} \A = -4 \beta = - 6 m(1+3\pi 
m^2/8g^2)^{-1/2}/\sqrt{\pi}g.
\label{thoofteq3} 
\end{equation}
We substitute Eq.~(\ref{thoofteq3}) in Eq.~(\ref{Moverm}) and obtain 
\beq
{M^2\over m^2} = 6\left[{(1+3\pi m^2/8g^2)^{1/2}-m/2\sqrt{\pi}g \over
(1+3\pi m^2/8g^2)^{1/2}-3m/4\sqrt{\pi}g}\right], \label{thoofteq4} 
\endeq
with the aid of $0 \leq m/g \le 2/ \sqrt {3 \pi}$. We calculate Eq. 
(\ref{thoofteq4}) and find 
\beq
2.45\,{m\over g}\le {M\over g} \le 2.56\,{m\over g}. 
\label{sugihararesult} 
\endeq
The average slope of the numerical result by Hornbostel et al. 
\cite{hornbostel}, after a correction, is
\begin{equation} M/g = 3.1\, m/g.
\label{hornbostelresult} 
\end{equation}
The SU(2) lattice gauge result of Hamer\cite{hamer} is 
\begin{equation} M/g = 3.2\, m/g. \label{hamerresult}
\end{equation}
A comparison of Eqs.~(\ref{simplestMm}), (\ref{thooftresult}), 
(\ref{sugihararesult}), (\ref{hornbostelresult}) and 
(\ref{hamerresult}), indicates that the ratios differ by about 30\%. 
\par
   Fig. 3 shows the comparison of the results of these models. It is 
apparent that all the models give values that are in general 
agreement. Thus, our simplest model $b_0(x)=1+\A x^2$ is enough to 
obtain the approximate lowest ground state mass.\par 

\section{SUMMARY AND CONCLUSIONS} We have studied the ratio of the 
ground
state mass $M$ to the quark mass $m$. The wavefunction $b_0(x)$ of 
(\ref{fulleq}) is expanded in a power series of even powers of $x$. 
We then matched equal powers of $x$ on both side of the equation 
(\ref{fulleq}). We have examined how each term of (\ref{power}) 
affects the final result. When 
$b_0(x)$ is set equal to 1, we find to order $x^2$ that $M=m=0$.\par
   Next, when $b_0(x) =1+Ax^2$, the $x^0$ and $x^2$ equations yield 
the result (\ref{simplestMm}), $M/g = \sqrt 6 m/g$. The substitution 
of $b_0(x) = \sin n\pi(x+1/2)$ in 't Hooft's equation gives for the 
ground state (n=1), $M/g =[\pi {\rm Si}(\pi)]^{1/2} m/g = 2.41\,m/g$, 
where ${\rm Si}(x)$ is the sine integral.\par 
   The simplest model of the wavefunction seems adequate to describe 
the features of the two dimensional light front description of 
mesons. The additional terms of the wavefunction $b_0(x)$ provide 
small corrections.\par 

\acknowledgements
One of us (O.A.) would like to thank Department of Physics, the Ohio 
State University for their hospitality, where this work was 
completed. 
\vfill\eject
\appendix
\section{DESCRIPTION OF THE FIT FOR $I_A(x)$, $I_B(x)$   AND $I_C(x)$}
In general, our functions $f_{1A,B,C,\cdots}$ and $f_{3A,B,C,\cdots}$ 
are parametrized in the form 
\beq
f_{ij}(y;r)=\sum_k^n b_{ij,k}(r){(-y)^{k/2}\over (\gamma_{ij} 
y+\delta_{ij})^{p_{ij}}(-y)^{q_{ij}}}, 
\endeq
with different values of the parameters as shown in Table 1. The 
$r$-dependence of $b_{ij,k}(r)$ is quadratic.\par 

\begin{table}[hbtp]
\renewcommand{\baselinestretch}{1}\tiny\normalsize
\def\mbx#1{\makebox[15mm]{#1}}
\caption{Values of parameters}
\label{table}
\begin{center}
\begin{tabular}{|c|c|c|c|c|c|} \hline
\mbx{}&\mbx{$n$}&\mbx{$\gamma$}&\mbx{$\delta$}&\mbx{$p$}&\mbx{$q$}
   \\ \hline\hline 
$f_{1A}$&    3&    0&    1&    1&    0\\ \hline
$f_{1B}$&    8&    1&    0&    0&    0\\ \hline
$f_{1C}$&    10&    1&    0&    0&    5\\ \hline
$f_{3A}$&    8&    1&    1&    2&    0\\ \hline
$f_{3B}$&    3&    0&    1&    1&    0\\ \hline
$f_{3C}$&    6&    1&    0&    0&    5\\ \hline
\end{tabular}
\end{center}
\renewcommand{\baselinestretch}{2}\tiny\normalsize
\end{table}

\centerline{\Large Figure Captions}
\bigskip
\def\p{\protect}
Fig. 1: The solutions to the integral equations for $M/g$ vs. $m/g$ 
in the present paper with the wavefunction expanded to order $x^2$, 
$x^4$, and  
$x^6$. The dotted line indicates the relation
$M/g=\sqrt{6}\,m/g$.
\vskip5mm
Fig. 2: To order $x^4$, the wavefunction is $b_0(x) = 1 + \A x^2 + 
\B x^4$. When $r$ and $\g$ are eliminated from the matches, the
parameters $\A$ and $\B$ are related. The real solutions are 
constrained as shown. 
\vskip5mm
Fig. 3: Results of the present work are compared to others 
calculations. Filled dots indicate the results of the present work, 
circles show the Hornbostel {\it et al.}'s results 
\p\cite{hornbostel}, open squares present the results of 't Hooft's 
large-$N$ expansion 
\p\cite{thooft},
and diamonds indicate Sugihara {\it et al}'s results 
\p\cite{sugihara}. The masses $M$ and $m$ are given in units of 
$\sqrt{g^2N_c/2\pi}$. 
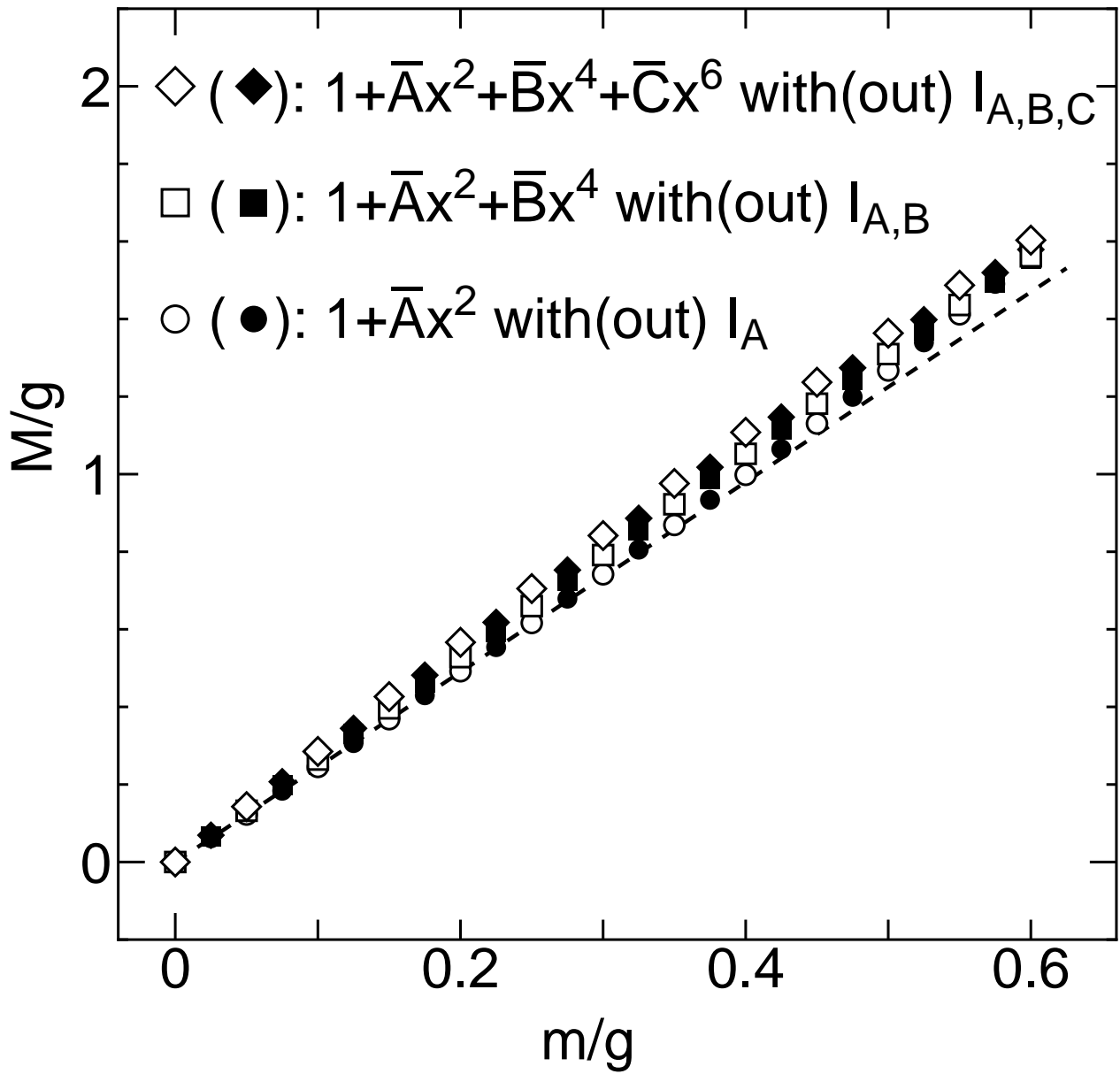
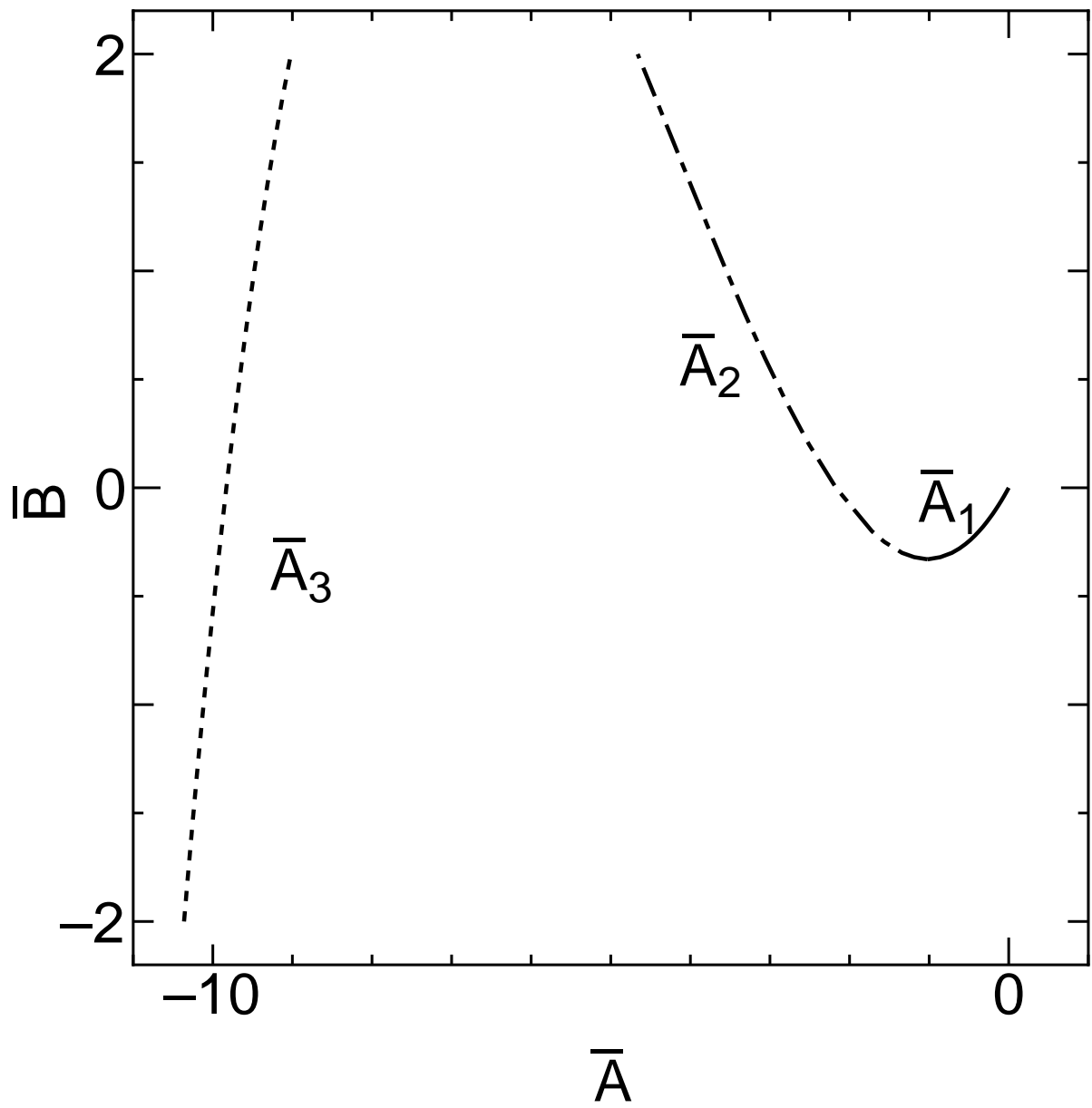
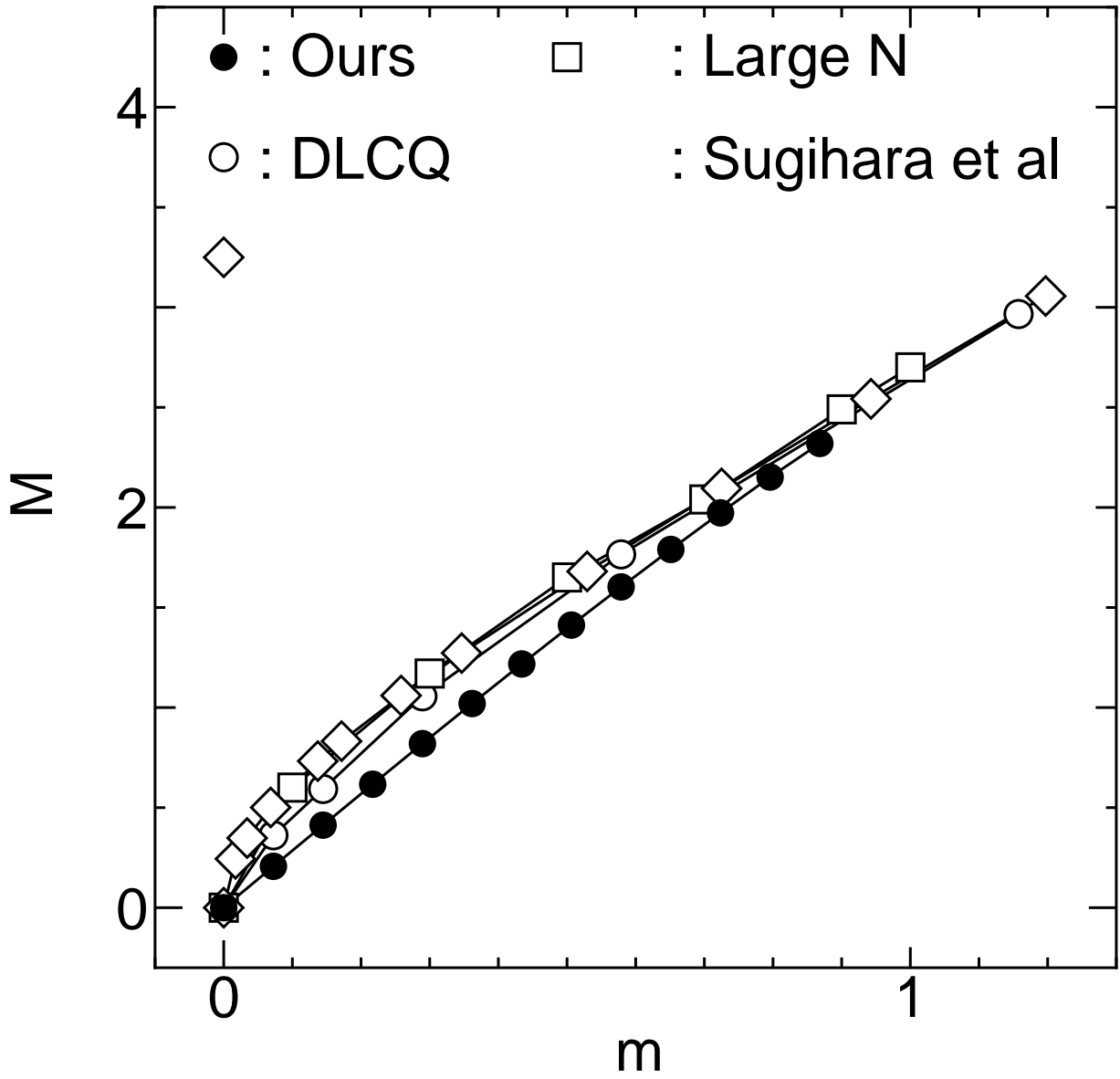
\end{document}